**Unraveling oscillations at ferro(para)magnetic and non-collinear antiferromagnetic interfaces**

Thiago Ferro[1], Luana Hildever[1], André José[2], and José Holanda[1,2,*]

[1]Programa de Pós-Graduação em Engenharia Física, Universidade Federal Rural de Pernambuco, 54518-430, Cabo de Santo Agostinho, Pernambuco, Brazil

[2]Unidade Acadêmica do Cabo de Santo Agostinho, Universidade Federal Rural de Pernambuco, 54518-430, Cabo de Santo Agostinho, Pernambuco, Brazil

* Corresponding author: joseholanda.silvajunior@ufrpe.br
*Orcid iD: https://orcid.org/0000-0002-8823-368X

**Abstract**

Here, we show that the ferro(para)magnetic and non-collinear antiferromagnetic interfaces contribute to oscillating signals observed in the magnetoresistance. We associate the effect with the fact that the spins on the surface of $IrMn_3$ produce instability in the surface magnetoresistance of the material, which is sensitive to the magnetic field. We carried out experiments using bilayers of $IrMn_3$ under permalloy (Py) and $IrMn_3$ under platinum (Pt). The oscillations were intensely evident at the Py/$IrMn_3$ interface and less explicit at the Pt/$IrMn_3$ interface. We carried out the experiments using pulsed current, with a square pulse width of 1μs and amplitudes of 20 μA to 20 mA. The oscillating signals are proportional to the crystallographic direction of the material, the ferromagnetism of the material adjacent to $IrMn_3$, and sensitive to the amplitude of the pulsed current. We believe that observation is a way of transmitting encoded information through magnetoresistance.

Spintronics (or spin electronics) studies the control of the flow from spin current in a device, using the charge of the electrons and its spin. Spintronics also is increasingly consolidating itself in antiferromagnetic materials, the so-called antiferromagnetic spintronics [1-5]. Antiferromagnets have better spin Hall properties than common non-magnetic materials. The spin Hall effects produce the coupling between charge currents and spin dynamics in magnetic systems. In this case, the symmetry-breaking mechanism can characterize the phases of matter. In an antiferromagnetic material (AFM), the Néel vector breaks the rotational symmetry of the paramagnetic state. The so-called easy axes, the particular directions of the spins in the crystal, are determined by the order parameters: Néel vector and spin-orbit coupling. These directions correspond to the lowest magnetocrystalline anisotropy energy (MAE). Certain AFMs, such as $Mn_2Au$, are good approximation isotropic Heisenberg magnets and can present in-plane MAE at levels from 10 µeV per formula unit [6]. An important non-collinear AFM with a giant MAE of 10 meV per formula unit is $IrMn_3$ [7]. A characteristic of $IrMn_3$ is that it has a spin Hall angle from 0.35 [8-10]. On the surface of thin films of antiferromagnets such as $IrMn_3$, the ferromagnetism of local uncompensated spins can induce symmetry of the antiferromagnet in an adjacent material [5]. The intensity of the interaction between $IrMn_3$ and the other material depends on the excitation at the surface. These are ingredients that influence the behavior of the non-magnetic spin Hall effect (NMSHE) [2, 3, 10-12] and the magnetic spin Hall effect (MSHE) [4-6, 9].

The conventional spin Hall effect or NMSHE is characterized by transverse spin currents that flow in non-magnetic conductors, generating an accumulation of spin near the sample surfaces [2-6, 9-13]. There are several possibilities to explore the phenomena involving the spin Hall effects [11, 12]. Non-magnetic materials are also well studied because current-induced spin polarization through spin-orbit interaction [2-6, 9-13] represents a potential application of spin transfer torque [13-16]. On the other hand, the MSHE has its interpretation in terms of the symmetries of well-defined linear response functions. In this way, the MSHE originates from a reactive counterpart of the dissipative spin response [4, 5, 9]. However, the magnetic order in the interfaces, non-collinear antiferromagnet/ferromagnet, and non-collinear antiferromagnet/paramagnet depends on the material symmetry and, consequently, on the direction of the spins. In this work, we report the experimental observation of oscillating signals in the magnetoresistance of $Py/IrMn_3$ and $Pt/IrMn_3$ bilayers. The oscillations in the $Py/IrMn_3$ bilayers are intense and evident, while those in the Pt/IrMn$_3$ bilayers are damped. **Fig. 1** shows a schematic of the three crystallographic configurations of the magnetoresistance MR = [(R(H)-R(H=0))/R(H=0)]x100% measurements carried out as a function of the magnetic field (H). The crystallographic direction is in emphasis because it is a way of manipulating the intensity of the observed effect. In **Fig. 1** the contacts of the wires of gold in samples were produced using the wire boner technique and considering the crystallographic directions. The distance between the external wires is 8 mm, between the internal is 4 mm and between them 2 mm.

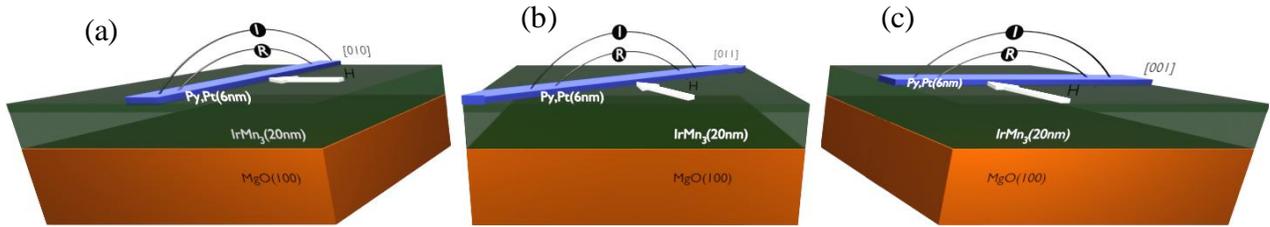

**Figure 1** (color on-line). Experimental scheme used to measure the magnetoresistance as a function of the magnetic field considering the three main crystallographic directions of the Py/IrMn$_3$ and Pt/IrMn$_3$ samples. **(a)** [010], **(b)** [011], and **(c)** [001]. Contacts of the wires of gold in samples were produced using the wire boner technique and considering the crystallographic directions. The distance between the external wires is 8 mm, between the internal is 4 mm and between them 2 mm.

All the films of IrMn$_3$ were epitaxial grown in a magnetron sputtering system with a base pressure of ~1 x 10$^{-7}$ torr with 3-mtorr argon and temperature of 840 K on the (100)MgO subtract. The Py and Pt layers were at a temperature of 470 K deposited. **Fig. 2 (a)** presents the X-ray diffraction in a sample of MgO/IrMn$_3$(20 nm)/Ti(2 nm), performed using a Philips Diffractometer with CuK-α radiation (λ = 1.54 Å). Considering the peak (200), we obtained the parameter *a* = 0.377 nm, which has the same value reported in refs. 4-6 and 9 [4-6, 9]. We realized the structural characterization of samples using a Transmission Electron Microscope (TEM) of 200 kV model JEOL 8100, as in ref. 17 [17]. We prepared the samples for TEM measurements using a Scanning Electron Microscope/Focused Ion Beam (SEM/FIB) model Zeiss NVision 40 with a cross-section. In **Fig. 2 (b)**, we show the transmission electron micrograph, which describes the different regions of each deposited material. It is possible to see each layer that composes the MgO/IrMn$_3$/Py sample. The sample interface is well defined as presented in other`s work [4-6, 9], just as it is possible to observe crystallized IrMn$_3$ by coupling the permalloy.

We performed the magnetic characterization using a Superconducting QUantum Interference Device (SQUID) of Quantum Designer, where we defined the exchange bias field as H$_{EX}$ = - (H$_1$ + H$_2$)/2, where H$_1$ and H$_2$ are the coercivities of the hysteresis loops. **Fig. 2 (c)** shows that the blocking temperature is above 300 K, where the exchange bias stabilized at room temperature, as in other works with the IrMn$_3$/Py bilayer [4, 6]. The exchange bias effect is a phenomenon that results from uncompensated spins at the interface of antiferromagnet/ferromagnet bilayers [4-6]. On most of the film surface, uncompensated antiferromagnetic spins are pinned in the direction of unidirectional anisotropy by antiferromagnetic domains [6, 7]. The difficulty of reorienting the AF domain defines the blocking temperature (T$_B$) [6].

The cooling from the heterostructure leads to the alignment of the preferred antiferromagnetic domain fixed below the T$_B$ and the anisotropy exchange about change in the direction of the magnetic field [4-7]. The thermal stability of the walls resulting in an antiferromagnetic domain determines the various characteristics of the exchange bias [5, 6, 9]. **Fig. 2 (c)** represents a decrease in the exchange bias field with the increase in temperature. At the temperature of 380 K, the exchange bias effect ceases to exist, defining the blocking temperature (T$_B$) for the IrMn$_3$/Py bilayer. The coupling of atoms Ni [18] to the atoms Mn [6, 19] at the interface defines the preferred direction by the antiferromagnetic domain in the IrMn$_3$ film as the sample is cooled [6, 9]. The described behavior justifies the sudden increase in the exchange bias field for temperatures below 100 K [6, 9, 20]. In this context, in **Fig. 2 (d)**, we show that the contacts of the wires of gold made using the wire boner technique are ohmic for all samples.

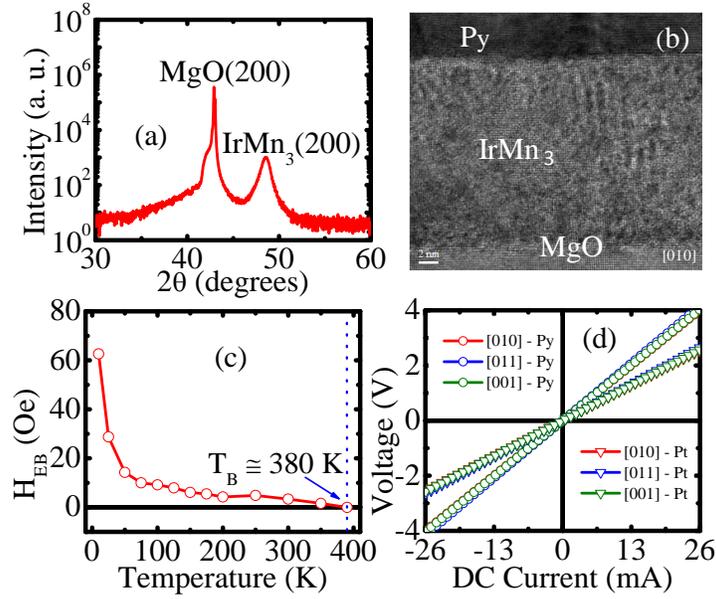

**Figure 2** (color on-line). **(a)** The X-ray diffraction measured on IrMn$_3$(20 nm) with a Ti cap layer with a thickness of 2 nm. **(b)** Show one micrograph of a typical sample MgO/IrMn$_3$(20 nm)/Py(6 nm). **(c)** It represents the change of the exchange bias field as a function of temperature, evidencing the blocking temperature (T$_B$ = 380 K). **(d)** Voltage curves as a function of DC electrical current show that the contacts of the wires of gold are ohmic.

An efficient way to investigate the antiferromagnetic interface is through electrical detection of the surface magnetic state in AFMs [21-23]. The spatial distribution of surface spins in non-collinear antiferromagnets can be very relevant since the surface distribution of these uncompensated spins is not the same as that of collinear antiferromagnets [14-16]. Another significant point is the difference between the ferromagnetic (or paramagnetic) and antiferromagnetic dynamics induced by the spin-polarized current [4-6, 9]. As is known, the value of the critical current that results in steady-state stability in an antiferromagnet can be greater than the corresponding value in a ferromagnet or paramagnet [4-6, 21]. The change of magnetoresistance as a function of the magnetic field for the different crystallographic directions shows a stable coupling in interfaces of the IrMn$_3$/Py (**Fig. 3 (a)** [010], **Fig. 3 (b)** [011], and **Fig. 3 (c)** [001]), and IrMn$_3$/Pt (**Fig. 3 (d)** [010], **Fig. 3 (e)** [011], and **Fig. 3 (f)** [001]) with the pulsed current of the 2mA from amplitude and 1 µs from width in a square pulse. In addition, we observed a uniform change throughout the magnetic field range. We realized all the magnetoresistance measurements to the room temperature using a Physical Property Measurement System (PPMS) of Quantum Designer.

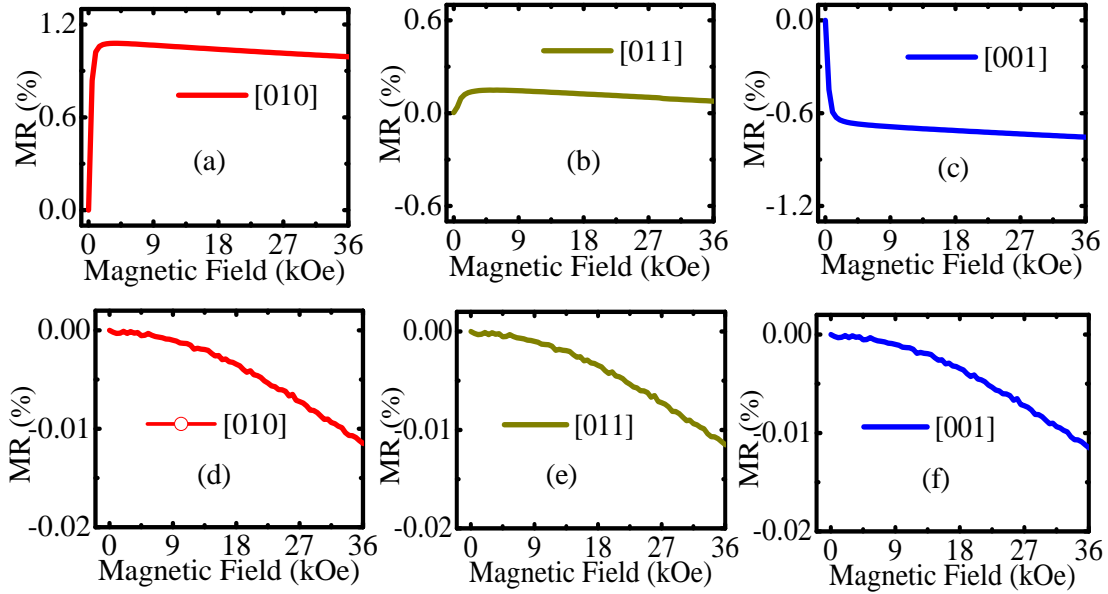

**Figure 3** (color on-line). Show the change from the magnetoresistance as a function of the magnetic field in three crystallographic directions. For the sample IrMn$_3$/Py: **(a)** [010], **(b)** [011], and **(c)** [001]. For the sample IrMn$_3$/Pt: **(d)** [010], **(e)** [011], and **(f)** [001]. We used a pulsed current of 2 mA from amplitude and 1 μs from width in a square pulse.

**Fig. 4** shows the oscillations obtained with a pulsed current from 20 mA amplitude with a square pulse width from 1μs. As discussed previously, the magnetoresistance effect can appear also in antiferromagnets, mainly due to uncompensated spins on the material surface. In this way, an electric current is a mechanism that can produce changes in the magnetoresistance of the material under a magnetic field. In **Figs. 4 (a)**, **(b)**, and **(c)**, it is possible to observe explicitly the oscillations. We also observed that the oscillating signals decrease even with continuous injection of spin currents towards the interface. For small magnetic fields, the effect results from the instability of the surface spins. As the magnetic field increases, more spins align, causing the oscillations to evolve with well-defined periods and phases.

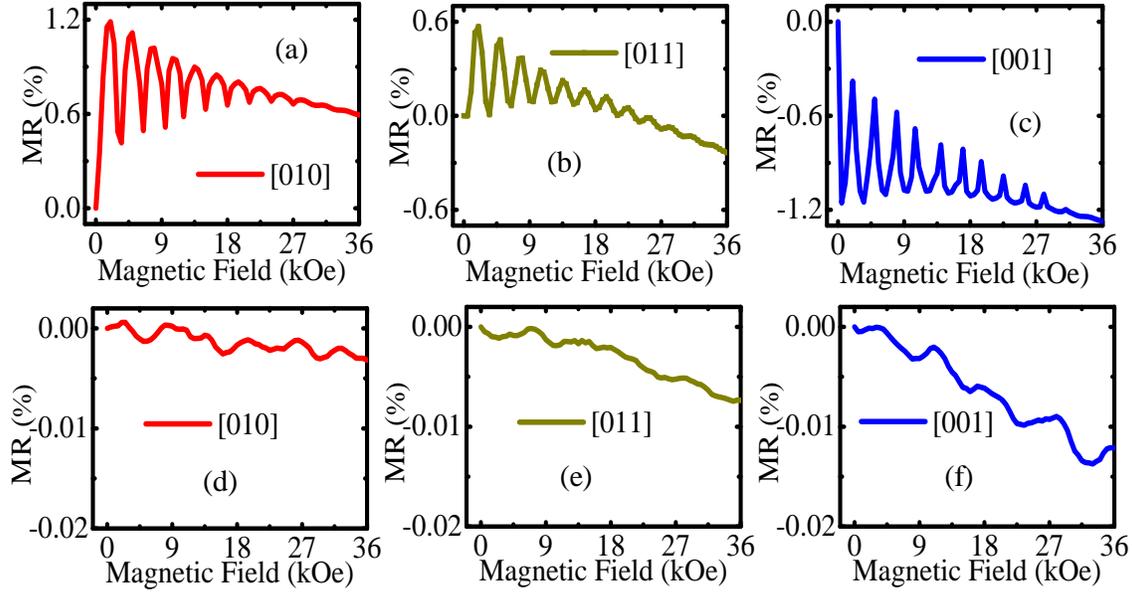

**Figure 4** (color on-line). Show the change from the magnetoresistance as a function of the magnetic field in three crystallographic directions. For the sample IrMn$_3$/Py: **(a)** [010], **(b)** [011], **(c)** [001]. For the sample IrMn$_3$/Pt: **(d)** [010], **(e)** [011], and **(f)** [001]. We used a pulsed current of 20 mA from amplitude and 1 μs from width in a square pulse.

For Pt as material adjacent to IrMn$_3$, the effect responsible for the observation from the oscillations at MR in the interface is the NMSHE. Oscillations are more damped because both materials have a large spin Hall angle. As is known, the platinum has a spin Hall angle of 5% and a diffusion length of ~ 5 nm [24], whereas the IrMn$_3$ has a spin Hall angle of 35% and a diffusion length of ~ 3nm [5, 6, 9]. Thus, spin accumulation on the interface is severely intense on both sides, but there is no material with well-defined polarization as in the case of permalloy. **Figs. 4 (d)**, **(e)**, and **(f)** show that the oscillations in a sample of the type (100)MgO/IrMn$_3$(20 nm)/Pt(6 nm) have characteristics of intense coupling between the materials regardless of the crystallographic direction. These oscillations have one effect relevant because one can control the spin flux depending on the value of the pulsed current and the intensity of the magnetic field. Then, it represents a new way of information encoding.

In summary, oscillations are observed and controlled mainly in bilayers of paramagnetic material under a ferromagnet or ferrimagnet [25, 26]. Here, we observed a new signal of oscillations across the magnetoresistance in Py/IrMn$_3$ and Pt/IrMn$_3$ bilayers. The main effects responsible for making the magnetoresistance oscillate are the NMSHE and the MSHE. The oscillating signals are proportional to the crystallographic direction of the material, the ferromagnetism of the material adjacent to IrMn$_3$, and sensitive to the amplitude of the pulsed current. The physical understanding of oscillations arises from the spin accumulation at the interface. In terms of basic and applied science, we believe that is an observation relevant.

**Acknowledgements**

This research was supported by Conselho Nacional de Desenvolvimento Científico e Tecnológico (CNPq), Coordenação de Aperfeiçoamento de Pessoal de Nível Superior (CAPES), Financiadora de Estudos e Projetos (FINEP), and Fundação de Amparo à Ciência e Tecnologia do Estado de Pernambuco (FACEPE). The authors are grateful to the researcher John E. Pearson from the Argonne National Laboratory and Prof. Eduardo Bedê from the Physics Department of the University Federal of Ceará for the valuable discussions on this work.

**Contributions**

T. F., L. H., and A. J. analyzed all the experimental measures and J. H. discussed, wrote and supervised the work.

**Conflict of interest**

The authors declare that they have no conflict of interest.